\documentclass[12pt]{article} 

\usepackage{ol2}
\usepackage[draft]{hyperref}
\usepackage{amsmath}

\begin{document}
\newcommand{\bq}{\begin{mathletters}}
\newcommand{\eq}{\end{mathletters}}
\newcommand{\beq}{\begin{eqnarray}}
\newcommand{\eeq}{\end{eqnarray}}
\newcommand{\beqq}{\begin{eqnarray*}}
\newcommand{\eeqq}{\end{eqnarray*}}

\newcommand{\st}[1]{{\mbox{${\mbox{\scriptsize #1}}$}}}  
\newcommand{\BM}[1]{\mbox{\boldmath $#1$}}              
\newcommand{\uv}[1]{\mbox{$\widehat{\mbox{\boldmath $#1$}}$}}   
\newcommand{\dy}[1]{\mbox{\boldmath $\overline{#1}$}}   
\newcommand{\sbtex}[1]{{\mbox{\scriptsize #1}}}   
\newcommand{\nab}{\mbox{\boldmath $\nabla$}}            
\newcommand{\CS}{\mbox{$\begin{array}{c}\cos \\ \sin \\ \end{array}$}}
\newcommand{\SC}{\mbox{$\begin{array}{c}\sin \\ \cos \\ \end{array}$}}

\newcommand{\rmi}{{\rm i}}
\newcommand{\er}{{\bf e}_{r}}
\newcommand{\ep}{{\bf e}_\varphi}
\newcommand{\et}{{\bf e}_\theta}
\newcommand{\hh}{{\bf H}}
\newcommand{\ee}{{\bf E}}
\newcommand{\ww}{{\bf W}}
\newcommand{\vv}{{\bf v}}
\newcommand{\rr}{{\bf r}}
\newcommand{\eee}{{\bf e}}
\newcommand{\uu}{{\bf u}}
\newcommand{\lom}{{\bf L}}

\newcommand{\rmq}{{\rm q}}
\newcommand{\rmt}{{\rm t}}
\newcommand{\co}{{\rm co}}
\newcommand{\cl}{{\rm cl}}

\newcommand{\epr}{\varepsilon_{r r}}
\newcommand{\mur}{\mu_{r r}}
\newcommand{\alr}{\alpha_{r r}}
\newcommand{\kar}{\kappa_{r r}}
\newcommand{\Om}{\Omega_a^r}
\newcommand{\ve}{\vec{e}}
\newcommand{\vw}{\vec{w}}
\newcommand{\va}{\vec{a}}


\title{An inverse method of designing cylindrical cloaks without knowing
coordinate transformation}


\author{C.-W. Qiu$^{1,2}$, A. Novitsky$^{3}$, and M. Soja$\check{c}$i$\acute{c}$$^{1}$}

\address{$^1$Research Laboratory of Electronics, Massachusetts Institute of
Technology, 77 Massachusetts Avenue, Cambridge, MA 02139, USA. \\
$^2$Department of Electrical and Computer Engineering, National
University of Singapore, Kent Ridge, Singapore 119620, Republic of
Singapore \\
$^3$Department of Theoretical Physics, Belarusian State University,
Nezavisimosti Avenue 4, 220050 Minsk, Belarus.\\
Corresponding authors. E-mail: cwq@mit.edu}

\begin{abstract}An inverse way to define the parameters of ideal cylindrical cloaks
is developed, in which the interconnection between the parameters is
revealed for the first time without knowing a specific coordinate
transformation. The required parameters are derived in terms of the
integral form of {\itshape cloaking generators}, which is very
general and allows us to examine the significance of the parametric
profiles. The validity of such inverse way and the invisibility
characteristics are presented in full-wave numerical simulation of
plane wave scattering by cloaked cylinders.
\end{abstract}



\noindent Designing invisibility cloaks has received increasing
interest from both engineering and scientific communities.
Coordinate transformation is commonly used to control
electromagnetic fields and render an object invisible to
electromagnetic radiation \cite{Pendry_sci,Leonhardt_sci2006}. This
approach is generalized from the cloaking in terms of the
conductivity \cite{Greenleaf_2003}, and further applied to acoustics
and electromagnetics \cite{Leonhardt_NJP,CTChan,Cummer}, which
provides the possibility of concealing not only passive but also
active objects within the interior \cite{ChenHS_2007,ZhangBL_2008}.

Given a virtual space with unit parameters, coordinate
transformation performs the transition from the non-scattering
virtual space to the realistic physical space, which in turn does
not interact with the incidence. It is due to the fact that
coordinate transformation method relies on the invariance of
Maxwell's equations throughout the spatial transformation. This
method can thus be applied to other classical waves (e.g., pressure
wave \cite{pressure} and surface liquid wave \cite{solid_wave})
excluding elastic wave \cite{Milton} provided that the symmetry
invariance of the corresponding wave equation is maintained under
the coordinate transformation. Therefore, traditional cloaks need to
be anisotropic and inhomogeneous. The anisotropy and the inhomogeity
of the cloak, as the deformation of the space, bend the wavefront
around the cloaked object and enable waves to emerge on the other
side along the propagation direction without any disturbance. Based
on this transformation technique, cylindrical cloaks with circular
\cite{cylinder}, elliptical \cite{ellipse} and arbitrary
cross-sections \cite{arb} have been studied, and practical attemps
to realize the cylindrical cloak have been made with promising
experimental results in microwave \cite{microwave} and optical \cite
{optical} regimes. Three-dimensional spherical electromagnetic cloak
has been studied analytically by Mie theory \cite{ChenHS_2007} for
first-order transformation and Von Neumann's method
\cite{high-order2} for higher-order transformation, respectively.
The realization of electromagnetic spherical cloaks by multilayered
isotropic coatings has been proposed \cite{Qiu_PRE09}.

Traditionally, in order to derive the anisotropic parameters for the
previous cloaks, one has to know the coordinate transformation
beforehand. On the contrary, in this letter, we present a novel
methodology to determine the required cloaking parameters for
cylindrical cloaks without knowing specific coordinate
transformations. Once any parameter is given, the rest parameters
can be derived in integral form associated with the introduced
cloaking generator. In this method, it is not even necessary to know
the exact profile of the first given parameter. The cloaking
generator together with the boundary condition, in fact, replace the
corresponding spatial transformation. Full-wave simulation results
are provided for verification. This work is an important step
forward in the search of desirable cylindrical cloaks via parametric
profiles.

For impedance matching purpose, both the relative permittivity and
permeability tensors of the cylindrical cloak are assumed to be
equal, i.e.,
$\dy{\varepsilon}(r)=\dy{\mu}(r)=\dy{\zeta}(r)=\zeta_r(r)\uv{r}\uv{r}+\zeta_\varphi(r)\uv{\varphi}\uv{\varphi}+\zeta_z(r)\uv{z}\uv{z}$.

First of all, we only know that a cylindrical cloak is designed by
compressing the virtual region $r'<b$ into the physical region
$a<r<b$ via an unknown prescribed transformation, where the prime
corresponds to the virtual space. The variables of $a$ and $b$ are
the inner and outer radius of the cylindrical cloak in the physical
space, respectively. It is assumed that the spatial compression is
only with respect to the radial direction, and thus the Jacobian
matrix is diagonal though we still have no information of the
specific form of the coordinate transformation. As a result, one can
obtain
\begin{equation}
\dy{\zeta}(r) =\left( \begin{array}{ccc} \zeta_r(r) & 0 & 0 \\ 0 &
\zeta_\varphi(r) & 0 \\ 0 & 0 & \zeta_z(r) \end{array} \right) =
\left( \begin{array}{ccc} \lambda_r/(\lambda_\varphi \lambda_z) & 0
& 0 \\ 0 & \lambda_\varphi/(\lambda_r \lambda_z) & 0
\\ 0 & 0 & \lambda_z/(\lambda_r \lambda_\varphi)
\end{array} \right),
\end{equation}
where
\begin{equation}
\lambda_r = \frac{d r}{d r'}, \qquad \lambda_\varphi = \frac{r}{r'},
\qquad \lambda_z = 1
\end{equation}
denote three principal stretches of the Jacobian matrix. Then three
equations can be derived:
\begin{equation}\label{lambdaEps_cyl}
\zeta_r(r) \zeta_\varphi(r) = 1,~~\zeta_r(r) \zeta_z(r) =
\frac{r^{\prime 2}}{r^2},~~\frac{d r'}{d r} = \sqrt{\zeta_\varphi(r)
\zeta_z(r)}.
\end{equation}

By manipulating Eq.~(\ref{lambdaEps_cyl}) and eliminating the term
$\zeta_\varphi(r)$, we derive the differential equation regarding
radial and transverse parameters
\begin{equation}\label{DErt1_cyl}
r \sqrt{\zeta_r(r) \zeta_z(r)} \frac{d \left[r \sqrt{\zeta_r(r)
\zeta_z(r)}\right]}{d r} = r \zeta_z(r).
\end{equation}

Integrating Eq.~(\ref{DErt1_cyl}), one has
\begin{equation}\label{epsRvsT_cyl}
r^2 \zeta_z(r) \zeta_r (r) = C + \int_a^r 2 r_1 \zeta_z(r_1) dr_1,
\end{equation}
where $C$ is the integration constant. Due to the spatial
compression ($r' = 0$ when $r=a$), it can be seen in
Eq.~(\ref{lambdaEps_cyl}) that $\zeta_r(r) \zeta_z(r)=0$, resulting
in $C=0$. Another requirement ($r' = b$ when $r=b$) leads to the
normalization condition for the $z$-component parameter
\begin{equation}
b^2 = \int_a^b 2 r_1 \zeta_z(r_1) d r_1,
\end{equation}
which plays an important role in finding cloaking parameters. Here,
we introduce {\itshape cloaking generator} $g(r)$ proportional to
$\zeta_z(r)$, i.e., $g(r)=C_0\zeta_z(r)$ where $C_0$ is an arbitrary
constant. Then $\zeta_z$ can be presented in the form
\begin{equation}
\zeta_z (r) =\frac{b^2g(r)}{2\int_a^b r_1g(r_1)dr_1}
\end{equation}

Radial and azimuthal parameters can be expressed as \begin{equation}
\zeta_r (r) = \frac{ 2 \int_a^r r_1 g(r_1) d r_1 }{r^2 g(r)};~~
\zeta_\varphi (r) = \frac{r^2 g(r)}{2 \int_a^r r_1 g(r_1) d
r_1}.\end{equation}

Only after all parameters are determined can corresponding
coordinate transformation for such cylindrical cloaks be found,
which is in contrast to the existing design approaches of
transformation based cloaks:
\begin{equation}\label{CT_cyl}
r' = b \sqrt{\frac{ \int_a^r r_1 g(r_1) d r_1 }{\int_a^b r_1 g(r_1)
d r_1}}.
\end{equation}

One may consider some interesting situations. When two
permittivities out of three are equal (i.e.,
$\zeta_r=\zeta_\varphi$, or $\zeta_z=\zeta_\varphi$, or
$\zeta_r=\zeta_z$), it can be shown that in each situation only the
trivial cloak ($r'=r$) is possible. Thus, ideal cylindrical cloak
has to be realized for three different permittivities:
$\varepsilon_r \neq \varepsilon_\varphi \neq \varepsilon_z$
\cite{microwave,QiuMin}. This is different from the case of
spherical cloaks with only two different parameters
\cite{Pendry_sci,Qiu_PRE09}.

\begin{table}
\caption{Cloaking parameters under different power cloak generators
for $\zeta_z(r)$ profile. The symbol $n$ is an arbitrarily positive
constant. We introduce $T_n(r)=(r-a)(a+r+nr)/(2+3n+n^2)$ to simplify
the expressions corresponding to the second power
profile.}\label{ta1}
\begin{center}
   \begin{tabular}{c|c|c|c|c}  \hline\hline
  $ \mbox{generator profile} $ & $ \zeta_z(r) $ & $ \zeta_r(r) $ & $ \zeta_\varphi(r) $ & $ \mbox{implied transformation} $
  \\ \hline
  $ \mbox{power}~g_1(r)=r^n $ &  $ \frac{b^2r^n(n+2)}{2(b^{n+2}-a^{n+2})} $ & $ \frac{2(r^{n+2}-a^{n+2})}{r^{n+2}(n+2)} $ & $\frac{r^{n+2}(n+2)}{2(r^{n+2}-a^{n+2})} $ & $ b\sqrt{\frac{r^{n+2}-a^{n+2}}{b^{n+2}-a^{n+2}}} $
\\ \hline
  $ \mbox{power}~g_2(r)=(r-a)^n $ &  $ \frac{b^2(r-a)^n}{2(b-a)^nT_n(b)} $ & $ \frac{2T_n(r)}{r^2} $ & $\frac{r^2}{2T_n(r)} $ & $ b\left(\frac{r-a}{b-a}\right)^{n/2} \sqrt{\frac{T_n(r)}{T_n(b)}}$
\\ \hline
   $ \mbox{power}~g_3(r)=(r-a)^n/r $ &  $ \frac{b^2(n+1)(r-a)^n}{2r(b-a)^{n+1}} $ & $ \frac{2(r-a)}{(n+1)r} $ & $\frac{(n+1)r}{2(r-a)} $ & $ b\left(\frac{r-a}{b-a}\right)^{(n+1)/2} $
\\
\hline\hline
\end{tabular}\\
{\small In $g_3(r)$, it is simply Pendry's classic cylindrical cloak
when $n=1$.}
\end{center}
\end{table}

To demonstrate the implementation of the cloaking generator and the
proposed inverse design method, we just consider and focus on three
different profiles of the power generators as given in Table~1.
Certainly, one may design other profiles for the cloaking generator
(e.g., parabolic, periodic, exponential, and even sinh profiles),
which is out of the scope of the current letter.

\begin{figure}
\centerline{\includegraphics[width=8cm]{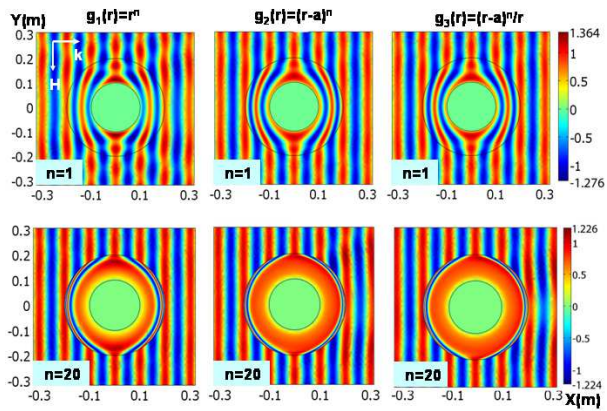}}
\caption{(Color online) The total electric field \mbox{Re[$E_z$]} on
the x-y plane when $n=1$ and $n=20$ for the power cloak generators.
The plane wave is propagating along x-axis and its electric field is
polarized along z-axis. The working frequency is 3GHz. The inner rod
is a PEC at the radius of $a=\lambda$, and the outer radius is
$b=2\lambda$.}
\end{figure}

When $n=0$, $g_1$ and $g_2$ power cloak generators are identical,
and their $\zeta_z$ components are constant. As an numerical
example, we present two sets of near-field patterns for the power
generators in Table~1, i.e., $n=1$ and $n=20$. In Fig.~1, it is
obvious that at $n=1$, $g_1$ power cloak exhibits larger disturbance
in the near-field pattern not only in the wavefront but also in the
variation of magnitude, and both $g_2$ and $g_3$ outperform $g_1$
under the same condition (the meshing of $n=1$ is the same for
$g_1$, $g_2$, and $g_3$) simulated in COMSOL Multiphysics. If the
meshing can be extremely fine, there will be no disturbance for all,
while it would require too much computer memory in simulation. This
non-ideality of the simulation conditions can be considered as a
non-ideal cloak consisting of a number of discrete cylindrical
layers. Also, at extremely high order (e.g., $n=20$), $g_1$
outperforms $g_2$ and $g_3$ since it results in smaller forward
scattering. The field distribution oscillates severely near the
outer radius in $g_2$ and $g_3$ power cloaks ($n=20$), implying that
the meshing has to be finer in this region. It is noted that the
field inside the cloak is more homogeneous at $n=20$ because at
larger $n$s the rays inside the cloaking shell will propagate very
close to the outer radius. From the transform point of view, if the
order of the power cloak generator becomes larger, more portions of
the original virtual space is transformed into the region close to
the outer boundary of the physical space. In theory, the cloaking
performance remains while the field distribution inside the cloak is
more squeezed into the outer radius. Thus, it needs much finer
meshing close to the outer radius, and the cases of $n=20$ all adopt
the same fine mesh. However, we cannot make the mesh extremely fine
due to the limitation of computer memory. Moreover, considering that
the order $n$ can be any arbitrary value, there may exist an optimal
range of $n$ in each power cloak in which the invisibility
performance will be further improved when the optimization is
applied to the disretized model \cite{Cummer09}, i.e., the
calculated total scattering width becomes smaller than that of
Pendry's classic one under the same computational criteria. In other
cloak designs other than power cloaks, it is possible to find such
optimal cases as well. Although these characteristics are of
technical importance, in our letter, we just focus on the proposed
inverse method which is useful to design cloak generators and
determine the optimal cloak in practice.

In this letter, we reported an inverse method to derive the cloaking
parameters for cylindrical cloaks which does not need to know the
required coordinate transformation first. It provides us larger
degree of freedom in the design of cylindrical cloaks in practice.
Instead of considering the coordinate transformation, we can
directly envisage the importance of the profile of just one
parameter, based on which the rest can be analytically and uniquely
determined. The numerical result confirms the validity of the
proposed approach, and also brings the community's attention to an
alternative design method for various cylindrical cloaks.

We thank Prof. Steven Johnson and Prof. John Joannopoulos for their
stimulating comments and discussion throughout the preparation of
this paper.


\begin{thebibliography}{99}


\bibitem{Pendry_sci} J. B. Pendry, D. Schurig, and D. R. Smith, ``Controlling electromagnetic fields," Science \textbf{312}, 1780 (2006).

\bibitem{Leonhardt_sci2006} U. Leonhardt, ``Optical conformal mapping," Science \textbf{312}, 1777-1780 (2006).

\bibitem{Greenleaf_2003} A. Greenleaf, M. Lassas, and G. Uhlmann, ``Anisotropic conductivities that cannot be detected by EIT," Physiol. Meas. \textbf{24},
413-419 (2003).

\bibitem{Leonhardt_NJP} U. Leonhardt, ``General relativity in electrical engineering," New J. Phys. \textbf{8}, 247 (2006).

\bibitem{CTChan} H. Chen and C. T. Chan, ``Acoustic cloaking in three dimensions using acoustic metamaterials," Appl. Phys. Lett. \textbf{91}, 183518 (2007).

\bibitem{Cummer} S. A. Cummer and D. Schurig, ``One path to acoustic cloaking," New J. Phys. \textbf{9}, 45 (2007).

\bibitem{ChenHS_2007} H. Chen, B. I. Wu, B. Zhang, and J. A. Kong, ``Electromagnetic wave interactions with a metamaterial cloak," Phys. Rev. Lett.
\textbf{99}, 063903 (2007).

\bibitem{ZhangBL_2008} B. Zhang, H. Chen, B. I. Wu, and J. A.
Kong, ``Extraordinary surface voltage effect in the invisibility
cloak with an active device inside," Phys. Rev. Lett. \textbf{100},
063904 (2008).

\bibitem{pressure} D. Torrent and J. Sanchez-Dehesa, ``Acoustic cloaking in two dimensions: a feasible approach," New J. Phys. \textbf{10}, 063015 (2008).

\bibitem{solid_wave} M. Farhat, S. Enoch, S. Guenneau, and A. B. Movchan, ``Broadband cylindrical acoustic cloak for linear surface waves in a fluid," Phys.
Rev. Lett. \textbf{101}, 134501 (2008).

\bibitem{Milton} G. W. Milton, M. Briane, and J. R. Willis, ``On cloaking for elasticity and physical equations with a transformation invariant form," New J. Phys. \textbf{8}, 248 (2006).

\bibitem{cylinder} D. Kwon and D. H. Werner, ``Two-dimensional eccentric elliptic electromagnetic cloaks," Appl. Phys. Lett. \textbf{92}, 013505 (2008).

\bibitem{ellipse} W. X. Jiang, T. J. Cui, G. X. Yu, X. Q. Lin, Q. Cheng, and J. Y. Chin, ``Arbitrarily elliptical¨Ccylindrical invisible cloaking," J. Phys. D: Appl. Phys. \textbf{41}, 085504 (2008).

\bibitem{arb} A. Nicolet, F. Zolla, and S. Guenneau, ``Electromagnetic analysis of cylindrical cloaks of an arbitrary cross section," Opt. Lett. \textbf{33}, 1584 (2008).

\bibitem{microwave} D. Schurig, J. J. Mock, B. J. Justice, S. A. Cummer, J. B. Pendry, A. F. Starr, and D. R. Smith, ``Metamaterial electromagnetic cloak at microwave frequencies," Science
\textbf{314}, 977 (2006).

\bibitem{optical} W. Cai, U. K. Chettiar, A. V. Kildishev, and V. M. Shalaev, ``Optical cloaking with metamaterials," Nat. Photonics \textbf{1},
063904 (2007).

\bibitem{high-order2} R. Weder, ``The boundary conditions for point transformed electromagnetic invisibility cloaks," J. Phys. A: Math. Theor. \textbf{41}, 065207
(2008).

\bibitem{QiuMin} M. Yan, Z. Ruan, and M. Qiu, ``Cylindrical invisibility cloak with simplified material parameters is inherently visible," Phys.
Rev. Lett. \textbf{99}, 233901 (2007).

\bibitem{Qiu_PRE09} C. W. Qiu, L. Hu, X. Xu, and Y. Feng, ``Spherical cloaking with homogeneous isotropic multilayered structures," Phys.
Rev. E \textbf{79}, 047602 (2009).

\bibitem{Cummer09} B. I. Popa and S. A. Cummer, ``Cloaking with optimized homogeneous anisotropic layers," Phys. Rev. A \textbf{79}, 023806 (2009).

\end{thebibliography}
\end{document}